\documentclass[11pt,a4paper]{article}
\usepackage{latexsym}
\usepackage{amsfonts}
\usepackage{amsmath}
\author{Diego Meschini\footnote{E-mail address: diego.meschini@phys.jyu.fi} \smallskip \\ \emph{Department of Physics, University of Jyv\"{a}skyl\"{a}}, \smallskip \\ \emph{PL 35 (YFL), FI-40014 Jyv\"{a}skyl\"{a}, Finland}.}
\title{Planck-scale physics: facts and beliefs}
\date{January 23, 2006}
\begin{document}
\maketitle
\begin{abstract}
The relevance of the Planck scale to a theory of quantum gravity has become a worryingly little examined assumption that goes unchallenged in the majority of research in this area. However, in all scientific honesty, the significance of Planck's natural units in a future physical theory of spacetime is only a plausible, yet by no means certain, assumption. The purpose of this article is to clearly separate fact from belief in this connection.

\smallskip


\end{abstract}

\section{Introduction} \label{Int}
An overview of the current confident status of so-called Planck-scale physics is enough to perplex. Research on quantum gravity has abundantly become synonymous with it, with the Planck length $l_P$, time $t_P$, and mass $m_P$ being unquestionably hailed as the scales of physical processes or things directly relevant to a theory of quantum gravity. And yet, in the last analysis, the only linking threads between Planck's natural units, $l_P$, $t_P$, and $m_P$, and quantum gravity ideas are (i) generally and predominantly, dimensional analysis---with all its vices and virtues---and (ii) specific only to a relative minority of research programmes, \emph{ad hoc} considerations about microscopic, Planck-sized black holes---a makeshift arena in which the effects of both quantum field theory and general relativity are to be relevant together.

Dimensional analysis is a surprisingly powerful method capable of providing great insight into physical situations without needing to work out or know the detailed principles underlying the problem in question. This (apparently) suits research on quantum gravity extremely well at present, since the physical mechanisms  to be involved in such a theory are unknown. However, dimensional analysis is not an all-powerful discipline: unless very judiciously used, the results it produces are not necessarily meaningful and, therefore, they should be interpreted with caution. So should, too, arguments concerning Planck-sized black holes considering, as we shall see, their ill-founded character.

In this article, we put forward several cautionary observations against the role uncritically bestowed currently on the Planck units as meaningful scales in a future physical theory of spacetime. One remark consists in questioning the extent of dimensional analysis' capabilities: although it possesses an almost miraculous ability to produce correct orders of magnitude for quantities from dimensionally appropriate combinations of physical constants (and physical variables\footnote{Due to its scope, in this article we shall only be concerned (with the exception of footnote 7) with physical quantities that depend on physical constants and not on physical variables. It should be noted, however, that dimensional analysis is normally applied to cases that involve the use of both physical constants and variables.}), it is not by itself a necessarily enlightening procedure. A conceptually deeper remark  has to do with the fact that, despite its name, a theory of quantum gravity need not be the outcome of ``quantum'' ($h$) and ``gravity'' ($G$, $c$): it may be required that one or more presently unknown natural constants be discovered and taken into account in order to picture the new physics correctly; or yet more unexpectedly, it may also turn out that quantum gravity has nothing to do with one or more of the above-mentioned constants.  

The meaning of these observations will be illustrated in the following with the help, where possible, of physical examples, through which we will cast doubt on the mostly undisputed significance of Planck-scale physics. We will argue that quantum gravity scholars, eager to embark on the details of their investigations, overlook the question of the likelihood of their assumptions regarding the Planck scale---thus creating seemingly indubitable facts out of merely plausible beliefs.

But before undertaking the critical analysis of these beliefs, let us first review the well-established \emph{facts} surrounding dimensional analysis and Planck's natural units.

\section{Facts} \label{Facts}
The Planck units were proposed for the first time by Max Planck (1899) over a century ago. His original intentions in the invention of these units were to provide a set of basic physical units that would be less arbitrary, i.e.\ less human-oriented, and more universally meaningful than metre, second, and kilogramme units.\footnote{For an account of the early history of the Planck units, see (Gorelik, 1992).} 

The Planck units can be calculated from dimensionally appropriate combinations of three universal natural constants, namely, Newton's gravitational constant $G$, Planck's constant $h$, and the speed of light $c$. The mathematical procedure through which Planck's units can be obtained is not guaranteed to succeed unless the reference set of constants $\{G, h, c\}$ is dimensionally independent with respect to the dimensions $M$ of mass, $L$ of length, and $T$ of time. This is to say that the $MLT$-dimensions of neither of the above constants should be expressible in terms of combinations of the other two. 

The  reference set $\{G,h,c\}$ will be dimensionally independent with respect to $M$, $L$, and $T$ if the system of equations 
\begin{equation} \left\{
\begin{matrix}
[G] & = & M^{-1}L^3T^{-2}\\
[h] & = & M^1L^2T^{-1}\\
[c] & = & M^0L^1T^{-1}
\end{matrix} \right.
\end{equation}
is invertible. Following Bridgman (1963, pp.~31--34), we take the logarithm of the equations to 
find
\begin{eqnarray} \left\{ \begin{matrix}
\ln([G])&=&-1\ln(M)+3\ln(L)-2\ln(T)\\
\ln([h])&=&1\ln(M)+2\ln(L)-1\ln(T)\\
\ln([c])&=&0\ln(M)+1\ln(L)-1\ln(T) \end{matrix} \right.,
\end{eqnarray}
which is now a linear system of equations in the logarithms of the variables of interest and can be 
solved applying Cramer's rule. For example, for $M$ we get
\begin{equation}
\ln(M)=\frac{1}{\Delta} \left|\begin{matrix}
\ln([G])&3&-2\\
\ln([h])&2&-1\\
\ln([c])&1&-1\\
\end{matrix}\right|,
\end{equation}
where 
\begin{equation} 
\Delta=\begin{vmatrix} 
-1 & 3 & -2 \\
1 & 2 &-1 \\
0 & 1 & -1 \\
\end{vmatrix}=2
\end{equation}
is the determinant of the system's coefficients. We find
\begin{equation} 
M=[G]^{-\frac{1}{\Delta}}[h]^\frac{1}{\Delta}[c]^{\frac{1}{\Delta}},
\end{equation}
and similarly for $L$ and $T$. This means that the system of equations has a solution only if $\Delta$ is different from zero, which in fact holds in this case. Granted the possibility to construct Planck's units, we proceed to do so.\footnote{Although Cramer's rule could be used to actually solve for $l_P$, $t_P$, and $m_P$, in the following we shall use a less abstract, more intuitive method.}

The Planck length $l_P$ results as follows. We define $l_P$ by the equation 
\begin{equation}
l_P=K_l G^\alpha h^\beta c^\gamma, 
\end{equation} 
where $K_l$ is a dimensionless constant. In order to find $\alpha$, $\beta$, and $\gamma$, we subsequently write its dimensional counterpart
\begin{equation}
[l_P]=L=[G]^\alpha [h]^\beta [c]^\gamma.
\end{equation} 
In terms of $M$, $L$, and $T$, we get
\begin{eqnarray}
L&=&(M^{-1}L^3T^{-2})^\alpha (ML^2T^{-1})^\beta (LT^{-1})^\gamma 
\label{length}
\end{eqnarray} 
Next we solve for $\alpha$, $\beta$, and $\gamma$ by comparing the left- and right-hand-sides of Eq.\ (\ref{length}) to get
$\alpha=\beta=1/2$, and $\gamma=-3/2$. Thus the Planck length is 
\begin{equation}
l_P=K_l \sqrt{\frac{Gh}{c^3}}=K_l 4.05\ 10^{-35}\ \mathrm{m}. 
\end{equation}
The other two Planck units can be calculated by means of a totally analogous procedure. They are 
\begin{equation}
t_P=K_t \sqrt{\frac{Gh}{c^5}}=K_t 1.35\ 10^{-43}\ \mathrm{s}
\end{equation}
and
\begin{equation}
m_P=K_m \sqrt{\frac{hc}{G}}=K_m 5.46\ 10^{-8}\  \mathrm{kg}. 
\end{equation}

The crucial question now is: is there any physical significance to these natural units beyond Planck's original intentions of providing a less human-oriented set of reference units for length, time, and mass? Quantum gravity research largely takes for granted a positive answer to this question, and confers on these units a completely different, and altogether loftier, role: Planck's units are to represent the physical scale of things relevant to a theory of quantum gravity, or at which processes relevant to such a theory occur.\footnote{See e.g.\ (Amelino-Camelia, astro-ph/0312014) for hypothetically possible roles of Planck's length.}

One may seek high and low for cogent justifications of this momentous claim, which is to be found in innumerable scientific publications; much as one might seek, one must always despairingly return to, by and large, the same explanation---when it is explicitly mentioned at all---dimensional analysis.\footnote{Interestingly, Butterfield and Isham (2000, p.~37) do account for the appearance of the Planck units, albeit in passing---and quite accurately indeed---as a ``simple dimensional argument.''} Is dimensional analysis such a trustworthy tool as to grant us definite information about unknown physics without needing to look into Nature's inner workings, or has our \emph{faith} in it become exaggerated? 

Two uncommonly cautious statements regarding the meaning of the Planck units are those of Y.J.~Ng's, and of J.C.~Baez's. Ng (2003) recognized that the Planck units are so extremely large or small relative to the scales we can explore today that ``it takes a certain amount of foolhardiness to even mention Planck-scale physics'' (p.~1\footnote{Page number refers to the online preprint gr-qc/0305019.}). 
Although this is a welcome observation, it appears to work only as a proviso, for Ng immediately moved on to a study of Planck-scale physics rather than to a criticism of it:  
\begin{quote}
But by extrapolating the well-known successes of quantum mechanics and general relativity in low energy, we believe one can still make predictions about certain phenomena involving Planck-scale physics, and check for consistency. (Ng, 2003, p.~1)
\end{quote} 

Also Baez (2000) made welcome critical observations against the hypothetical relevance of the Planck length in a theory of quantum gravity. Firstly, he mentioned that the dimensionless factor (here denoted $K_l$) might in fact turn out to be very large or very small, which means that the order of magnitude of the Planck length as is normally understood (i.e.\ with $K_l=1$) need not be meaningful at all. A moment's reflection shows that this problem does not really threaten the expected order of magnitude of the Planck units. An overview of physical formulas suggests that, in general, the dimensionless constant $K$ appearing in them tends to be very large or very small (e.g.\ $|K|\geq 10^2$ or $|K|\leq 10^{-2}$) only when the physical constants enter the equations to high powers.\footnote{For example, consider a black-body and ask what might be its radiation energy density $\rho_E$. Since the system in question involves a gas of photons, one proposes
\begin{equation}
\rho_E=K_\rho h^\alpha c^\beta (k_B\tau)^\gamma,    
\end{equation}
where $K_\rho$ is a dimensionless constant and $\tau$ is the absolute temperature of the black body. The application of dimensional analysis gives the result 
\begin{equation}
\rho_E=K_\rho \frac{(k_B\tau)^4}{h^3c^3}=K_\rho 4.64\ 10^{-18}\tau^4 \frac{\mathrm{J}}{\mathrm{K}^4\mathrm{m}^3}=7.57\ 10^{-16}\tau^4\ \frac{\mathrm{J}}{\mathrm{K}^4\mathrm{m}^3}  \qquad (K_\rho=163). \label{energydensity}
\end{equation}  
Therefore, the result yielded by dimensional analysis is two orders of magnitude lower than the correct value of $\rho_E$, which represents a meaningful difference for the physics of a black-body. This discrepancy, however, could have been expected since it can be traced back to the constants $h$, $c$, and $k_B$ entering the equation to high powers. See (Bridgman, 1963, pp.~88, 95) for related views.} In this respect, the situation appears quite safe for the Planck units, excepting perhaps the Planck time in whose expression $c$ enters to the power $-5/2$.

More interestingly, Baez also recognized that ``a theory of quantum gravity might involve physical constants other than $c$, $G$, and $\hbar$.'' (p.~180). This is indeed one important issue regarding the significance of Planck-scale physics---or lack thereof---which will be analyzed in Section \ref{Beliefs}. However, notwithstanding his cautionary observations to the effect that``we cannot prove that the Planck length is significant for quantum gravity,'' Baez also chose to try and ``glean some wisdom from pondering the constants $c$, $G$, and $\hbar$'' (p.~180).  

What these criticisms intend to point out is the curious fact that, within the widespread uncritical acceptance of the relevance of Planck-scale physics, even those who noticed the possible shortcomings of dimensional analysis in connection with quantum gravity decided to continue to pursue their studies in this direction. Where do the charms of dimensional analysis lie, then? 

``[D]imensional analysis is a by-way of physics that seldom fails to fascinate even the hardened practitioner,'' said Isaacson and Isaacson (1975, p.~vii). Indeed, examples could be multiplied at will for the view that dimensional analysis is a trustworthy, almost magical, tool. Consider, for example, a hydrogen atom and ask what is the relevant physical scale of its radius $a$ and binding energy $E$. Dimensional analysis can readily provide an answer \emph{apparently} without much knowledge of the physics of atoms, only if one is capable of estimating correctly which natural constants are relevant to the problem.\footnote{The manner of this estimation will be considered in Section \ref{Appraisal}. It is assumed here that one at least knows what an atom \emph{is}, namely, a microscopic aggregate of lighter negative charges and heavier positive charges of equal strength, and where only the \emph{details} of their mutual interaction remain unknown. Without this kind of previous knowledge, in effect assuring us that we have a quantum-electrodynamic system in our hands, the application of dimensional analysis becomes \emph{blind}. See (Bridgman, 1963, pp.~50--51).} Armed with previous physical experience, one notices that only quantum mechanics and the electrodynamics of the electron are involved, so that the constants to consider must be $h$, the electron's charge $e$, the electron's mass $m_e$, and the permittivity of vacuum $\epsilon_0$. Equipped with these constants, one proposes
\begin{eqnarray}
a&=&K_a h^\alpha e^\beta m_e^\gamma \epsilon_0^\delta\\
E&=&K_E h^\lambda e^\mu m_e^\nu \epsilon_0^\xi,    
\end{eqnarray}
where $K_a$ and $K_E$ are dimensionless constants. A procedure analogous to the one above for the Planck length, gives the results 
\begin{eqnarray}
a&=&K_a \frac{\epsilon_0 h^2}{m_e e^2}=K_a 1.66 \ \mathrm{\AA}=0.529 \ \mathrm{\AA} \qquad \left(K_a=\frac{1}{\pi}\right) \label{radius}\\
E&=&K_E\frac{m_e e^4}{\epsilon_0^2 h^2}=K_E 108\ \mathrm{eV}=-13.60\ \mathrm{eV} \qquad \left(K_E=-\frac{1}{8}\right), \label{energy}
\end{eqnarray}  
where the values of $K_a$ and $K_E$ do not result from this analysis but are obtained from detailed physical calculations; these values are given here for the purpose of assessing the accuracy of dimensional analysis. As we can see, meaningful values of the order of magnitude of the hydrogen atom's characteristic radius and binding energy can thus be found through the sheer power of dimensional analysis. We believe that it is on a largely unstated argument along these lines that quantum gravity researchers' claims regarding the relevance of the Planck units rest. 

Regrettably, the physical world cannot be probed confidently by means of this tool alone, since it has its shortcomings. As we will see in the next section, these are that dimensional analysis may altogether fail to inform us what physical quantity the obtained order of magnitude refers to; and more severely, much can go amiss in the initial process of estimating which natural constants must be consequential and which not. Therefore, any claims venturing beyond the limits imposed by these shortcomings, although plausible, must be based on \emph{faith}, \emph{hope}, or \emph{belief}, and should not in consequence be held as straightforward facts.
       
\section{Beliefs} \label{Beliefs}
Our first objection against a securely established significance of the Planck units in quantum gravity lies in the existence of cases in which dimensional analysis yields the order of magnitude of no clear physical thing or process at all. 

To illustrate this point, take the Compton effect. A photon collides with an electron, after which both particles are scattered. Assume---as is the case in quantum gravity---that the detailed physics behind the effect is unknown, and use dimensional analysis to predict the order of magnitude of the length of any thing or process involved in the effect. Since quantum mechanics and the dynamics of an electron and a photon are involved here, one assumes (quite correctly) that the natural constants to be considered are $h$, $c$, and $m_e$.\footnote{Like before, it is here essential to be assured that the physical system under study is quantum-electrodynamic.} To find this Compton length $l_C$, we set 
\begin{equation}
l_C=K_Ch^\alpha c^\beta m_e^\gamma,
\end{equation}
where $K_C$ is a dimensionless constant, to find 
\begin{equation}
l_C=K_C\frac{h}{m_e c}=0.0243\ \mathrm{\AA} \qquad (K_C=1).
\end{equation}
What is now $l_C$ the length of? In the words of Ohanian (1989), ``in spite of its name, this is not the wavelength of anything'' (p.~1039). 

Ohanian's statement is controversial and requires a qualification. In fact, one immediate meaning of the Compton length can be obtained from the physical equation in which it appears, namely, $\Delta \lambda=l_C [1-\cos(\theta)]$, where $\lambda$ is the photon's wavelength and $\theta$ is the scattering angle. It is straightforward to see that $l_C$ must be the change of wavelength of a photon scattered \emph{at right angles} with the target electron.
In this manner, a meaning can be found to any constant appearing in a physical equation by looking at the special case when the functional dependence is set to unity. Perhaps this explains Ohanian's denial of any real meaning to $l_C$, since via this method meaning can be attached to any constant whatsoever.\footnote{Baez (2000) explained another physical meaning of the Compton length as the distance at which quantum field theory becomes necessary for understanding the behaviour of a particle of mass $m$, since ``determining the position of a particle of mass $m$ to within one Compton wavelength requires enough energy to create another particle of that mass'' (p.~179). Here again, a new meaning can be found for $l_C$ based on already existing, detailed knowledge provided by quantum field theory.}

In this sense, the meaning of constants appearing in physical equations can be learnt from their very appearance in them. However, this takes all charm away from dimensional analysis itself, since such equations are not provided by it but become available only after well-understood physical theories containing them are known; i.e.\ after the content of the equations is related to actual observations. In consequence, the discovery of the Planck units' meaning in a theory of quantum gravity depends on such equations being first discovered and interpreted against a background of actual observations.\footnote{Another possibility is that the meaning of physical constants could be obtained from simpler, theory-independent methods through which to measure them directly; however, that we have such a more direct method is certainly not true today of the Planck units. Neither is such an extreme operationalistic stance required to give meaning to them; it suffices to have a (Planck-scale) theory of quantum gravity making \emph{some} observable predictions. But at least this much is necessary in order to make quantum gravity a meaningful \emph{physical} theory, and thus disentangle the claims of this currently speculative and volatile field of research.} And yet, what guarantees that the Planck units \emph{will} meaningfully appear in such a future theory? Only the very \emph{definition}---supported, as we shall see, by a hasty guess---of quantum gravity as a theory involving the constants $G$, $h$, and $c$. This brings us to our second objection. 

A more severe criticism of the widespread, unquestioned reign of the Planck units in quantum gravity research results from conceptual considerations about quantum gravity itself, including an exemplifying look at the history of physics. 

As mentioned above, it is a largely unchallenged assumption that quantum gravity will involve precisely what its \emph{name} makes reference to---the quantum ($h$) and gravity ($G,c$)---and (i) nothing \emph{more} or (ii) nothing \emph{less}. The first alternative presupposes that in a future theory of spacetime, and any observations related to it, the combination of \emph{already known} physics---and nothing else---will prove to be significant.\footnote{Guessing on available knowledge is only natural and in itself unobjectionable. As expressed earlier on, our criticism is rather directed at the uncritical \emph{attitude} with which these guesses are usually made and proclaimed.} As correct as it might turn out to be, this is too restrictive a conjecture for it excludes the possibility of the need for \emph{truly new} physics (cf.\ Baez's observations). In particular, is our current physical knowledge so complete and final as to disregard the possibility of the existence of relevant natural constants yet undiscovered?\footnote{Curiously, the \emph{mainstream} assumption of only $G$, $h$, and $c$ as relevant to quantum gravity pays little heed to other readily available aspects of the Planck scale. For example, why not \emph{routinely} consider Boltzmann's constant $k_B$ and the permittivity of vacuum $\epsilon_0$ (or how about other suitable constants?) to get Planck's temperature $\tau_P=K_\tau (hc^5/Gk_B^2)^{1/2}=K_\tau 3.55\ 10^{32}$ K, and Planck's charge $q_P=(hc\epsilon_0)^{1/2}=1.33\ 10^{-18}$ C as relevant to quantum gravity as well? Just because only length, time, and mass appeal to our more primitive, mechanical intuition? Or is it perhaps because the notions of temperature and charge are not included in the \emph{label} ``quantum gravity''? For approaches considering extended Planck scales, see e.g.\ (Cooperstock \& Faraoni, 2003; Major \& Setter, 2001).} The second alternative takes for granted that \emph{at least} all three constants $G$, $h$, and $c$ must play a role in quantum gravity. Although this is a seemingly sensible expectation, it need not hold true either, for a theory of quantum gravity may also be understood in less conventional ways. For example, not as a quantum-mechanical theory of (general-relativistic) gravity but as a quantum-mechanical theory of \emph{empty spacetime}, as we explain below.    
 
In order to present the import of the first view more vividly, we offer an illustration from the history of physics. Reconsider the hydrogen atom problem before Bohr's solution was known and \emph{before} Planck's solution to the black-body problem was ever given, i.e.\ before any knowledge of the quantum, and therefore of $h$, was available.\footnote{This demand is, of course, anachronistic since $h$ \emph{was} known to Bohr; this notwithstanding the hypothetical situation we propose serves as the basis for a completely plausible argument. It is only a historical accident that the problem of the collapse of the atom was first confronted with knowledge of the quantum, since the latter was discovered while studying black-body radiation, a problem independent of the atom. However, since the black-body problem does not lend itself to the analysis we have in mind, we prefer to take the example of the hydrogen atom, even if we must consider it anachronistically.} With no solid previous experience about atomic physics, the atom would have been considered an electromagnetic-mechanical system, and dimensional analysis could have been (unjustifiably) used to find the order of magnitude of physical quantities significant to the problem. Confronted with this situation, a nineteenth-century quantum gravity physicist may have decidedly assumed that no new constants need play a part in yet unknown physical phenomena, proceeding as follows. 

Since an electron and a (much more massive) proton of equal charge are concerned, the constants to be considered must be $e$, $m_e$, $\epsilon_0$, and the permeability of vacuum $\mu_0$---and nothing else. This is of course wrong, but only to the modern scientist who enjoys the benefit of hindsight. The inclusion of $\mu_0$ is not at all unreasonable since it could have been suspected that non-negligible magnetic effects played a part, too. 

In order to find the order of magnitude of the hydrogen atom's radius $a$ and binding energy $E$, one sets
\begin{eqnarray}
a&=&K'_a e^\alpha m_e^\beta \epsilon_0^\gamma \mu_0^\delta\\
E&=&K'_E e^\lambda m_e^\mu \epsilon_0^\nu \mu_0^\xi.
\end{eqnarray}
Proceeding as before, one obtains
\begin{eqnarray}
a&=&K'_a \frac{e^2\mu_0}{m_e}=K'_a 3.54\ 10^{-4}\ \mathrm{\AA}\\
E&=&K'_E \frac{m_e}{\epsilon_0\mu_0}=K'_E 5.11\ 10^5\ \mathrm{eV}.
\end{eqnarray}
\emph{These expressions and estimates for $a$ and $E$ are completely wrong} (see Eqs.\ (\ref{radius}) and (\ref{energy}) for the correct results). This comes as no surprise to today's physicist, who can tell at a glance that the constants assumed to be meaningful are wrong: $\mu_0$ should not be there at all, and $h$ has not been taken into account. How can then today's quantum gravity physicists ignore the possibility that they, too, might be missing one or more yet unknown constants stemming from genuinely new physics?

The second, less conventional view expressed above, namely, that quantum gravity need not be understood as a quantum-mechanical theory of (general-relativistic) gravity, is supported by the following reasoning. In view of the repeated difficulties and uncertainties encountered so far in attempts to uncover gravity's quantum-mechanical aspects, one may wonder whether the issue might not rather be whether spacetime \emph{beyond its metric field}---i.e.\ empty spacetime as characterized by its bare points---may have quantum-mechanical aspects. In particular and as we enquired in (Meschini, Lehto, \& Piilonen, 2005; Meschini \& Lehto, gr-qc/0506068), may the consideration of quantum theory reveal any physical reality (in the form of observables) that empty spacetime possesses but which classical general relativity denies to them (cf.\ hole argument)? From this perspective, gravity's possible quantum-mechanical features are not an issue and, if one stands beyond the metric field---itself the bearer of spacetime's gravitational features ($G$) and causal structure ($c$)---the constants $G$ and $c$ have no reason to arise in the theory.

\subsection{Quantum gravity from Planck-scale black holes?}
As advanced in the introduction, there exists one specific approach to quantum gravity that does not 
rely on dimensional analysis in order to reproduce Planck's length and mass. Baez (2000, pp.~179--180), Thiemann (2003, pp.~8--9\footnote{Page numbers refer to the online preprint gr-qc/0210094.}), and to some extent Saslow (1998) have all independently put forward the essence of the argument in question from different perspectives. The gist of the idea is that in order for a wave packet of mass $m$ to be spread no more than one Compton length $l_C=h/mc$, its energy spectrum $\Delta E$ must be spread no less than $mc^2$, i.e.\ enough energy to create---according to quantum field theory---a particle similar to itself. One the other hand, in order for the same wave packet to create a non-negligible gravitational field (strong enough to interact with itself), it must be spread no more than one Schwarzschild radius $l_S=2Gm/c^2$, i.e.\ be concentrated enough to become---according to general relativity---a black hole. Both effects are deemed to become important together when $l_C=l_S$, i.e.\ when the particle becomes a black hole of mass $(hc/2G)^{1/2}=m_P/\sqrt{2}$ and radius $(2Gh/c^3)^{1/2}=\sqrt{2}l_P$---a so-called 
Planck black hole,\footnote{The appearance of the factor $\sqrt{2}$ in this derivation is not normally mentioned; Baez (2000), e.g., deliberately ignores it. It is also noteworthy that Planck's time is connected with this argument only according to predictions from quantum field theory in curved spacetime, namely, that a Planck black hole is unstable with a lifetime of $t_P\approx 10^{-43}$ s. However, this theory, in which spacetime is taken to be classically curved but not ``quantized,'' is commonly referred to as a semiclassical theory (of quantum gravity) and paradoxically distrusted to hold at the Planck scale.} the paradigmatic denizen of the quantum-gravitational world.  

Is this a sound argument or is it yet another \emph{ad hoc}, \emph{ad lib} armchair exercise devised only to quench our thirst for tangible, ``quantum-gravitational things''? Regardless of the complicated and physically controversial question whether such black holes actually exist, let us concentrate here only on the conceptual aspect of the issue. Our criticism has two parts. 

Firstly, it is straightforward to see that this reasoning to the effect that Planck black holes must be paradigmatic of quantum gravity assumes the previously criticized view that such a theory must arise out of a combination of the quantum and gravity. It is really no surprise to find that, on this supposition, the Planck scale \emph{does} result even without the intervention of dimensional analysis; now instead of simply combining universal constants together, we combine theories which contain them. What the concept of a Planck black hole proves, then, is something we already knew, namely, that the Planck scale must be relevant to a theory which glues the pieces of our current knowledge together.

Secondly, the manner in which the argument is construed appears as notoriously haphazard as the procedures of dimensional analysis. In effect, general relativity is first assumed nonchalantly in order to implement the idea of a Planck black hole, but the latter concept is immediately used to argue against the validity of general relativity---and therefore of black holes!---at the Planck scale, where it is quantum gravity that has the upper hand. (Or can there be black holes without general relativity? Does the mythical, unknown yet true, theory of quantum gravity allow for them?) Therefore, the reasoning seems to be conceptually flawed. Perhaps, one of the greatest shocks that quantum gravity has in store for us is the tragicomical revelation that Planck black holes cannot exist at all.

The welcome view is sometimes expressed that the likelihood of  this notion is uncertain. For example, Thiemann shares (for his own reasons) our last point that the notion of Planck black holes must be conceptually flawed:
\begin{quote}
[A Planck black hole] is again [at] an energy regime at which quantum gravity must be important and these qualitative pictures must be fundamentally wrong\ldots (Thiemann, 2003, p.~9)
\end{quote}
Interestingly enough, Thiemann takes this \emph{negative} argument \emph{optimistically}, while we can only take it pessimistically---as a bad omen.

Given the conceptually uncertain nature and relevance of this idea, we conclude that one should take it with a pinch of salt, and not anywhere more seriously than considerations arising from dimensional analysis.

\section{Appraisal} \label{Appraisal}
The state of affairs reviewed in Section \ref{Beliefs} leads to serious doubts regarding the applicability of dimensional analysis. How can we ever be sure to trust any results obtained by means of this method? Or as Bridgman (1963) jocularly put it in his brilliant, little book: ``We are afraid that\ldots we will get the incorrect answer, and not know it until a Quebec bridge falls down'' (p.~8). It is therefore worthwhile asking: where, in the last analysis, did we go wrong? 

In this respect, Bridgman enlightened us by explaining that a trustworthy application of dimensional analysis requires the benefit of \emph{extended experience} with the physical situation we are confronted with as well as \emph{careful thought}:

\begin{quote}
We shall thus ultimately be able to satisfy our critic of the correctness of our procedure, but to do so requires a considerable background of physical experience, and the exercise of a discreet judgment. The untutored savage in the bushes would probably not be able to apply the methods of dimensional analysis to\ldots [a] problem and obtain results which would satisfy us. (Bridgman, 1963, p.~5)
\end{quote}
And further:
\begin{quote}
The problem [about what constants and variables are relevant] cannot be solved by the philosopher in his armchair, but the knowledge involved was gathered only by someone at some time soiling his hands with direct contact. (Bridgman, 1963, pp.~11--12)
\end{quote}

In this light, a conscientious application of dimensional analysis to the affairs of quantum gravity is hopeless. There does not exist so far any realm of physical experience pertaining to interactions of gravitation and quantum mechanics, and therefore there is no previous physical knowledge available and no basis whatsoever on which to base our judgement. Thus, the quantum gravity researcher proclaiming the relevance of the Planck scale resembles Bridgman's savage in the bushes or philosopher in the armchair, not because of being untutored or not wanting to soil his hands, but because there are no observational means available to obtain any substantial information about the situation of interest.   

Bridgman also dismissed any meaning to be found in Planck's units considering the manner in which they arise:
\begin{quote}
The attempt is sometimes made to go farther and see some absolute significance in the size of the [Planck] units thus determined, looking on them as in some way characteristic of a mechanism which is involved in the constants entering the definition. 

The mere fact that the dimensional formulas of the three constants used was such as to allow a determination of the new units in the way proposed seems to be the only fact of significance here, and this cannot be of much significance, because the chances are that any combination of three dimensional constants chosen at random would allow the same procedure. Until some essential \emph{connection} is discovered between the \emph{mechanisms} which are accountable for the gravitational constant, the velocity of light, and the quantum, it would seem that no significance whatever should be attached to the particular size of the units defined in this way, beyond the fact that the size of such units is determined by phenomena of universal occurrence. (Bridgman, 1963, p.~101) [Italics added.]
\end{quote}

Gorelik (1992), and von Borzeszkowski and Treder (1988) in fact criticized Bridgman's early remarks above on account that, indeed, the connection has now been found under the form of quantum gravity. A moment's reflection shows that this is a deceptive argument. One cannot legitimately contend that the new connection found is quantum gravity, and therefore that Planck's units have meaning because, with no phenomenological effects to support it, there is no physical substance to so-called quantum gravity besides (mostly Planck-scale based) theoretical speculations. Therefore, the argument is at best unfounded and at worst circular.

To sum up, the relevance of Planck-scale physics to quantum gravity rests on several uncritical assumptions, which we have here cast doubt upon. Most importantly, there is the question of whether quantum gravity can be straightforwardly decreed to be the simple combination of the quantum and gravity. This excluding attitude does not appear wise after realizing, on the one hand, that black-body physics did not turn out to be the combination of thermodynamics and relativity, nor atomic physics the combination of electromagnetism and mechanics; and on the other hand, that quantum gravity can also be interpreted in a less literal sense---for what's in a name after all?---and regarded to include, e.g., spacetime features beyond gravity. Dimensional and \emph{ad hoc} analyses only yield meaningful results when the appropriate physics is known to start with. To presume that we did not know enough in this respect a hundred years ago but that we do now is to presume too much. 

Further, we argued that the physical meaning of the Planck units could only be known after the successful equations of the theory which assumes them---quantum gravity---were known. To achieve this, however, the recognition and observation of some phenomenological effects genuinely related to spacetime are essential. Without them, there can be no trustworthy guide to oversee the construction of, and give meaning to, the ``equations of quantum gravity,'' and one must then resort to wild guesses. Ultimately, herein lies the problem of dimensional and \emph{ad hoc} analyses as applied to quantum gravity. The fact nevertheless remains that no genuine spacetime observables, neither classical nor quantum-mechanical, have as yet ever been identified. It is only through the discovery and observation of such new phenomena telling us what quantum gravity is \emph{about} that we will gradually unravel the question concerning what quantum gravity \emph{is}.  

In view of this uncertainty surrounding Planck's natural units, we believe it would be more appropriate to the honesty and prudence that typically characterize the scientific enterprise, not to abstain from their study if not so wished, but simply to express their relevance to quantum gravity as a humble belief, and not as an established fact, as is regrettably today's widespread practice.

\section*{Acknowledgements}
The author is grateful to his supervisor, Dr.\ Markku Lehto, for his thought-provoking guidance through the labyrinths of quantum gravity research.

\section*{References}
\begin{enumerate}
\item Baez, J.~C.\ (2000). Higher-dimensional algebra and Planck-scale physics. In C.~Callender, \& N.~Huggett (Eds.), \emph{Physics meets philosophy at the Planck scale} (pp.~177--195). Cambridge: Cambridge University Press. 

\item Bridgman, P.~W.\ (1963). \emph{Dimensional analysis}. New Haven, London: Yale University Press. (Original revised edition published 1931.)

\item Butterfield, J., \& Isham, C.\ (2000). Spacetime and the
philosophical challenge of quantum gravity. In C.~Callender, \& N.~Huggett (Eds.), \emph{Physics meets philosophy at the Planck scale} (pp.~33--89). Cambridge: Cambridge University Press. 

\item Cooperstock, F., \& Faraoni, V.\ (2003). The new Planck scale: Quantized spin and charge coupled to gravity.  \emph{International Journal of Modern Physics D}, \emph{12}, 1657--1662.  

\item Gorelik, G.\ (1992). First steps of quantum gravity and the Planck values. In J.~Eisenstaedt, \& A.~J.~Kox.\ (Eds.), \emph{Studies in the history of general relativity}, Einstein Studies, Vol.~3 (pp.~364--379). Boston: Birkh\"{a}user. 

\item Isaacson, E.~de St Q., \& Isaacson, M.~de St Q.\ (1975). \emph{Dimensional methods in engineering and physics}. London: Arnold.

\item Major, S.~A., \& Setter, K.~L.\ (2001). On the universality of the entropy-area relation. \emph{Classical and Quantum Gravity}, \emph{18}, 5293--5298.

\item Meschini, D., Lehto, M., \& Piilonen, J.\ (2005). Geometry, pregeometry and beyond. \emph{Studies in History and Philosophy of Modern Physics}, \emph{36}, 435--464.
 
\item Ng, Y.~J.\ (2003). Selected topics in Planck-scale physics. \emph{Modern Physics Letters A}, \emph{18}, 1073--1097. 

\item Ohanian, H.~C.\ (1989). \emph{Physics} (2nd ed.). New York: Norton and Co.

\item Planck, M.\ (1899). \"{U}ber irreversible Strahlungsvorg\"{a}nge. \emph{Sitzungsberichte der Preussischen Akademie der Wissenschaften}, \emph{5}, 440--480.

\item Saslow, W.~M.\ (1998). A physical interpretation of the Planck length. \emph{European Journal of Physics}, \emph{19}, 313.

\item Thiemann, T.\ (2003). Lectures on loop quantum gravity. \emph{Lecture Notes in Physics}, \emph{631}, 41--135. 

\item von Borzeszkowski, H., \& Treder, H.\ (1988). \emph{The meaning of quantum gravity}. Dordrecht: Reidel.

\end{enumerate}

\end{document}